# Spin-orbit torque in Pt/CoNiCo/Pt symmetric devices

Meiyin Yang[1†], Kaiming Cai[1†], Hailang Ju[2], Kevin William Edmonds[3], Guang Yang[4], Shuai Liu[2], Baohe Li[2], Bao Zhang[1], Yu Sheng[1], ShouguoWang[4], Yang Ji[1] & Kaiyou Wang[1*]

1. SKLSM, Institute of Semiconductors, CAS, P. O. Box 912, Beijing 100083, People's Republic of China
2. Department of Physics, School of Sciences, Beijing Technology and Business University, Beijing 100048, China
3. School of Physics and Astronomy, University of Nottingham, Nottingham NG7 2RD, United Kingdom
4. Department of Materials Physics and Chemistry, University of Science and Technology Beijing, 100083, China
† These authors contributed equally to this work.
*Correspondence and requests for materials should be addressed to K. W.( kywang@semi.ac.cn)

Current induced magnetization switching by spin-orbit torques offers an energy-efficient means of writing information in heavy metal/ferromagnet (FM) multilayer systems. The relative contributions of field-like torques and damping-like torques to the magnetization switching induced by the electrical current are still under debate. Here, we describe a device based on a symmetric Pt/FM/Pt structure, in which we demonstrate a strong damping-like torque from the spin Hall effect and unmeasurable field-like torque from Rashba effect. The spin-orbit effective fields due to the spin Hall effect were investigated quantitatively and were found to be consistent with the switching effective fields after accounting for the switching current reduction due to thermal fluctuations from the current pulse. A non-linear dependence of deterministic switching of average $M_z$ on the in-plane magnetic field was revealed, which could be explained and understood by micromagnetic simulation.

Strong interests have been focused on the electrical control of the magnetic moment in spintronic devices due to its promising application in low-power consumption and high-speed processing logic and memory[1]. Recently, the torques induced by in-plane current in heavy metal (HM)/FM/oxide samples with strong spin-orbit coupling have been shown to offer a very efficient way to manipulate the moment of ferromagnets[2,3], induce domain wall motion[4,5] and cause persistent magnetic oscillation[6]. A favorable characteristic of these switching mechanisms is the simplicity of the film structure, which does not require a magnetic polarization layer. The physical mechanism underlying the magnetization switching by spin-orbit torque consists of at least two components, the spin Hall effect and the Rashba effect. The damping-like torque generated by spin currents from the spin Hall effect was reported to explain the magnetization switching[2]. One indication was that the sign of the magnetization switching reversed between samples with Pt and Ta underlayers due to their opposite sign of the spin Hall angle[7], which also proved that the spin-orbit torque is a function of the spin Hall angle[8]. On the other hand, the field-like torque induced by the Rashba effect due to structural asymmetry was also claimed to be the dominant contribution to current-induced switching of perpendicular magnets[3,9,10]. The main reason for this opinion was that experimentally the switching effective field was far larger than the theoretically predicted spin Hall induced effective field. The phase diagram of the deterministic switching by spin Hall effect under in-plane magnetic field simulated for a single domain model did not agree well with the experiment[11]. In addition, the thickness of Ta in Ta/CoFeB/MgO was found to influence the sign of the field-like torque and even the damping-like torque[12], suggesting the Rashba effect may play a significant role.

In the HM/FM/oxide systems, it is difficult to quantitatively separate the switching contributions from spin Hall and Rashba effects directly, since they coexist in this structure and generate both damping-like torques of the form $a\boldsymbol{m}\times\boldsymbol{\sigma}\times\boldsymbol{m}$ (DL, or Slonczewski-like) and field-like torques of the form $b\boldsymbol{m}\times\boldsymbol{\sigma}$ (FL) with different directions[13,14], where $\boldsymbol{m}$ and $\boldsymbol{\sigma}$ are direction and vectors of ferromagnet's magnetization and spin of the current, respectively. In order to quantitatively understand the spin Hall effect contribution to the magnetization switching, a symmetric film structure (HM/FM/HM) was fabricated to minimize the Rashba effect. However, the capping HM layer has the opposite influence of spin Hall effect to that of the bottom HM layer, so that the overall spin Hall effect could be eliminated experimentally[15]. Here, we utilize a symmetric structure of the form, Pt/FM/Pt, to study the spin Hall effect induced magnetization switching, which as far as possible minimizes the Rashba effect experimentally but maintains the spin Hall effect due to the different current paths through the structure. In order to eliminate the spin Hall effect from the upper Pt capping layer, we designed the device structure to make sure that the FM layer can mostly sense the spin current from the bottom Pt layer, which arises mainly from the spin Hall effect. After accounting for the current-induced thermal effect, the effective magnetic field due to the spin Hall effect is in good agreement with the measured switching field. The deterministic switching under a range of in-plane fields, $H_x$, directed parallel to the current direction, was also investigated. The value of $M_z$ after sweeping the current through a cycle dramatically decreased with lowering the value of $H_x$. This result is shown to be in agreement with LLG micromagnetic simulations. Our work will explain apparent inconsistencies between the experiments and the theory of the spin-orbit torque from the spin Hall effect.

## Results

**Device structure and current distributions.** The structure of the device is shown in Fig. 1a. The FM layer and the upper Pt layer are patterned into a 3 $\mu$m diameter disk, which sits above a Hall cross fabricated from the lower Pt layer. The FM layer consists of a 0.8 nm Co/Ni/Co trilayer, henceforth referred to as CoNiCo. The fabrication process is described in the Method section. The hysteresis loop of the anomalous Hall effect (AHE) for the 3 $\mu$m magnet with the field applied in the out-of-plane direction is presented in Fig. 1b. The square hysteresis loop shows that the CoNiCo dot has strong perpendicular anisotropy with a switching field of 520 Oe. The magnetic properties of the unpatterned film were measured by ferromagnetic resonance (FMR). The results, shown in Supplementary S1, reveal an anisotropy constant $K_u$ of $8\times10^6$ erg per cm$^{-3}$ and damping constant of ~0.05. The electrical current was applied along the bottom Pt layer, leading to spin accumulation at its top and bottom surfaces. The current distribution at the position of the magnetic dot in the multilayer structure was calculated using the current continuity equation and Ohm's law (detailed calculation can be found in Supplementary S2). The calculated current density distribution through a two-dimensional slice of the structure is plotted in Fig. 1c. The red arrows indicate the direction and magnitude of the current density. The current in the CoNiCo and upper Pt layers follows an arc, and at its very edge flows in the vertical direction. The in-plane components of the current density distributions are shown in Fig. 1d. The averaged current density of the upper Pt, CoNiCo, and the lower Pt layer are approximately in the ratio 1:1.5:17.5 from the numerical calculation. The spin accumulation from the upper Pt layer would be less than 3% compared with the bottom Pt, considering their current densities and thickness differences. Thus we could ignore the spin Hall effect from the upper Pt layer based on this calculated result.

**Effective fields generated by the current.** The two interfaces of HM/FM/HM samples are not quite identical[16] even if the two HM layers are the same metals. To find out whether the two interfaces in the symmetric structure of our device produce a net Rashba effect, harmonic measurements were conducted to quantitatively determine the field-like effective fields in the Pt/CoNiCo/Pt structure, which mostly result from the Rashba effect.

Second harmonic measurements provide a very powerful probe of the field-like torque and damping-like torque in spin-orbit coupled systems[17]. We conducted the measurement by applying an a.c. current with frequency $f$ to the lower Pt electrodes and measuring its first and second harmonic Hall voltages simultaneously using two lock-in amplifiers. The a.c. current generates alternating effective fields, oscillating the magnetization around its equilibrium position and raising the second harmonic voltage $V_{2f}$. The $V_{2f}$ Hall voltage contains information on the magnetization oscillation angle due to the current-induced effective field. The damping-like and field-like effective fields can be calculated with the following equation when the magnetization stays near the $z$ axis:

$$H_{DL(FL)} = -2\frac{H_{L(T)} \pm 2\xi H_{T(L)}}{1-4\xi^2}, \tag{1}$$

where $\xi$ is the ratio of planar Hall resistance and anomalous Hall resistance ($R_{PHE}/R_{AHE}$), $H_T$ and $H_L$ are defined as $H_{L,\pm} = \left(\partial V_{2f,L\pm}/\partial H_x\right)/\left(\partial^2 V_{f,L\pm}/\partial H_x^2\right)$ and $H_{T,\pm} = \left(\partial V_{2f,L\pm}/\partial H_y\right)/\left(\partial^2 V_{f,T\pm}/\partial H_y^2\right)$, respectively. The $\pm$ sign indicates the magnetization pointing up and down. The longitudinal ($H_x$) and transverse ($H_y$) magnetic fields are the applied external magnetic fields parallel and perpendicular to the current direction in the film plane during the measurements. The AHE and PHE were measured using the first harmonic voltage with the field switching out of plane and rotating in the plane, respectively (Supplementary S3). Normally the ratio $\xi$ is very small, 0.13 in our case, due to the smaller value of the PHE. Thus, based on the above equations, the $H_{DL}$ ($H_{FL}$) mostly depends on $H_x$ ($H_y$) when the external field is applied along the current direction (transverse to the current direction).

A diagram of the measurement is shown in Fig. 2a, where the a.c. current was applied along the *x* axis with external field along *x* ($H_x$) and *y* ($H_y$) axis. The first order and second order Hall voltages with magnetic field applied in both the *x* and *y* directions are shown in Fig. 2b and 2d. Clear and smooth curves were observed with $H_x$ applied under different a.c. current amplitudes. However, there is no obvious observed signal when sweeping the magnetic field along the *y* axis under the same current value. The signal under $H_y$ is below the measurement resolution (15 nV) and smaller than 1% of the signal under $H_x$, indicating that $H_{FL}$ in the designed symmetric structure is below the limit of the measurements. Since the $H_{FL}$ arises mostly from the Rashba effect, we conclude that the Rashba effect is largely eliminated in our device structure. This phenomenon is consistent with previous studies in which the symmetric structure Pt/Co/Pt did not have the second harmonic signal while large $H_{FL}$ was observed in Pt/Co/AlO$_x$, suggesting little Rashba effect in the Pt/Co/Pt sample[10]. Using equation (1), the current-induced effective fields were calculated and presented in Fig. 2c. No noticeable $V_{2f}$ signal under $H_y$ field was obtained as mentioned above.

The $H_{DL}$ can be obtained and increases linearly with the current amplitude. The damping-like effective field induced by the current was estimated to be 25 Oe per $10^7$ Acm$^{-2}$.

**Spin Hall effect contribution to the current-induced switching.** To understand how much the damping-like effective field induced by the spin Hall effect contributes to the current-induced switching, we investigated the current-induced switching in the device by applying a 100 ms current pulse to the Pt layer under different perpendicular external field. The critical switching current density ($\pm J_c$) linearly decreases with the magnitude of the external magnetic field, as is shown in Fig. 3a. The effective field generated by the electrical current was estimated to be 60 Oe per $10^7$ Acm$^{-2}$ from the slope. The magnitude of the effective field obtained by the critical current switching measurements is more than twice that obtained from the harmonic measurements (25 Oe per $10^7$ Acm$^{-2}$). This result is similar to a previous study where the spin Hall induced effective field was reported to be smaller than the switching effective field[9]. The differences often are used as the evidence for current-induced switching driven by the Rashba effect. However, apart from the Rashba effect, the thermal effect associated with the electrical current also contributes to the magnetization switching. The relationship between the perpendicular external fields and the critical switching current density without thermal excitation can be described by the Slonczewski's model[18]: $J_{c_0} = A\alpha M_s t(H_k - 4\pi M_S - H)/g(\theta)P$, where $A$ is a constant of $3\times 10^8$ A per Oe per emu, $M_S$ is the saturation magnetization, $\alpha$ is the Gilbert damping, $t$ is the thickness of CoNiCo multilayer, $P$ is the spin polarization of the current, $H_k$ is the perpendicular anisotropy field and $g(\theta) = [-4 + (1+P)^3(3 + S_1 \cdot S_2/4P^{3/2})]^{-1}$ where $S_1$ and $S_2$ are the current spin direction and the ferromagnetic layer spin direction, respectively. However, in our case the switching pulse length is 100 ms where the thermal contribution to the magnetization switching must be considered. The critical switching current with increasing pulse length considering the effect of thermal fluctuations follows the equation[19, 20]: $I_c(\tau) = I_{c_0}[1 - K_B T/K_u V \ln(\tau/\tau_0)]$, where $I_{c_0}$ is the critical switching current density when the pulse length is 1 ns, $K_B$ is the Boltzmann constant, $K_u$ is the uniaxial anisotropy, $T$ is the temperature, $V$ is the volume of the magnetic layer, $\tau$ is the length of the current pulse and $\tau_0$ is 1 ns. The experimental data of $I_c$ under the perpendicular field of 500 Oe versus $\tau$ are presented in Fig. 3b and its fitting gives us the $K_u V / K_B T$ parameter of 35.

Given that the spin accumulates at the Pt top and bottom interfaces due to the current in the Pt layer, the spin current was in the vertical direction and diffused into the CoNiCo layer. Thus, the spin current density depends on the spin Hall angle of the Pt. After considering both the thermal effect and spin Hall effect in our system, the critical switching current density could be rewritten as:

$$J_c = \frac{A\alpha M_s t [1 - \ln(\tau/\tau_0) K_B T/K_u V]}{g(\theta)[1 - \text{sech}(d/\lambda_{sf})] P(J_s(d=\infty)/J_e)} (H_k - 4\pi M_s - H), \qquad (2)$$

where $d$ is the thickness of Pt layer, and $\lambda_{sf}$ is the diffusion length of Pt (~2 nm) (ref. 20). $P$ is chosen to be 1, since the spin Hall effect generates the pure spin current. Applying equation (2) to fit the experiment data in Fig. 3a, the fitted Hall angle of 0.03 for the bottom Pt (3 nm) layer was obtained. The bulk spin Hall angle $J_s(d=\infty)/J_e$ was 0.052, which is in the range of published values[21,22,23]. The switching current density decreased by almost half after considering the reduction of the energy barrier due to thermal activation. The effective field excluding the thermal effect is estimated to be around 30 Oe per $10^7$ Acm$^{-2}$ from the critical current density measurements with perpendicular magnetic fields, which is quite consistent with the results from the harmonic measurements. Our results show that the thermal effect contribution, which was seldom considered previously, nearly equals the effect of spin-orbit torques in our measurement.

We also compared our results with the Pt/CoFeB/AlO$_x$ sample from the previous report[24]. There, $H_{DL}$ and $H_{FL}$ were found to be around 55 Oe per $10^7$ Acm$^{-2}$ and 176

Oe per $10^7$ Acm$^{-2}$, respectively, using the harmonic measurements. The Pt layer of the sample in ref. 24 is thicker than the one used in this paper and gives a larger spin Hall angle (0.06), indicating a larger spin current generation. After normalizing the spin Hall angle to the same value used in this paper (0.03), the calculated $H_{DL}$ is about 22.5 Oe per $10^7$ Acm$^{-2}$, almost the same as the result in this work, 25 Oe per $10^7$ Acm$^{-2}$. Surprisingly, the effective field obtained from the critical current switching measurements, 52 Oe per $10^7$ Acm$^{-2}$, is also similar to our results. However, the electrical current switching efficiency is not reduced even though the $H_{FL}$ is eliminated for the symmetric structure Pt/CoNiCo/Pt.

The $H_{FL}$ effective fields which originate mostly from the Rashba effect are directed along the $y$ axis. This may be the reason it does not much influence the switching current densities, for much larger in-plane fields (6,000 Oe) are required to fully align the magnetization to the field plane in Fig. 2c. We also applied external fields in the plane with different angles between the field direction and the current direction, and found no reduction of the switching current density (Supplementary S5). However, for the spin Hall effective field (mostly, $H_{DL}$), there will be a perpendicular component during the magnetization switching, which greatly enhances its switching efficiency, as the critical current densities are sensitive to the perpendicular magnetic fields (Fig. 3a). Using Slonczewski's model and the harmonic measurements, we can conclude that in the Pt/CoNiCo/Pt symmetric device, the spin Hall effect assisted by thermal fluctuations from the current pulse is sufficient to switch the magnets.

**Current induced switching under $H_x$ fields.** In order to give further insight into the deterministic switching by spin-orbit torque, the current switching measurements were performed over a range of in-plane external fields. The out-of-plane magnetization of a perpendicular magnet can be determined by applying an in-plane external field during the current switching[2,7]. When the in-plane field is not sufficiently large, the $M_z$ cannot be fully switched. The damping-like torque from the spin-Hall effect can give us a reasonable explanation of this phenomenon, in which case the perpendicular effective fields are formed by the damping-like torque with the symmetry broken by the $H_x$ (ref. 7).

The hysteresis loops of the anomalous Hall resistance as a function of the current with the in-plane field of ±400 Oe are shown in Fig. 4a. The magnetization was switched from $+M_z$ to $-M_z$ with the +400 Oe external field when sweeping the current from negative to positive, and switched back from $-M_z$ to $+M_z$ when sweeping the current from positive to negative. With a -400 Oe external field applied, the opposite switching behavior is observed. Similar measurements were performed with different in-plane magnetic fields applied, ranging from 60 to 950 Oe (Fig. 4b). The critical current density is not sensitive to the magnitude of the in-plane magnetic field. The averaged $M_z$ at zero current corresponding to different in-plane fields are revealed from the anomalous Hall resistance loops. The $M_z$ first increases and then decreases with increasing magnitude of the in-plane magnetic field, which was plotted in Fig. 4d. The maximum Hall resistance is detected at in-plane magnetic field of 400 Oe, and it dramatically reduces to zero when the in-plane field was gradually removed. For in-plane fields larger than 400 Oe, the equilibrium magnetization of the CoNiCo dot deviated significantly from the perpendicular direction, resulting in the shorter $M$ projection along $z$ axis and the smaller $R_h$.

The critical in-plane magnetic fields along $x$ direction from experiments was found to be around five to ten times smaller than the values predicted by the single domain model[11]. To analyze the averaged $M_z$ under in-plane external field quantitatively, we simulated the switching process calculation using a commercial LLG micromagnetic simulator[25]. We simulated the current-induced switching under in-plane fields < 400 Oe where the magnetization breaks into domains. The model consists of a 0.8 nm thin cylinder with a diameter of 100 nm, spin polarized electrons along $y$ axis, a spin current flowing perpendicular to the cylinder for 5 ns and with applied magnetic fields along $x$ axis. The initial magnetization was set to (0, 0, 1), as the material has a perpendicular anisotropy. The external magnetic field sustained after the end of the current pulse. The time-dependence of the spatially averaged magnetization along *x, y, z* was recorded. Only Slonczewski-like torque was included in this simulation, since the field-like torque was too small to be detected in our experiment. In Fig. 4c we present the time-dependence of the averaged *M* projected along *z* axis under in-plane external field of 100 Oe and 400 Oe, respectively. Under the field of 100 Oe, the $M_z/M_s$ drops sharply for the first 1 ns and then stays at a relative stable value of -0.18, and slightly decreased and stabilized at -0.26 after the current pulse ended. For a larger external field at 400 Oe, the $M_z/M_s$ gradually decreased to a lower value of -0.71 and precessed to -1 after the current pulse, meaning the magnetization is fully switched to –*z* axis (Supplementary Fig. S7). Fig. 4d compares the $M_z/M_s$ of the experiment data, the simulated result at the end of the 5 ns current pulse, and the stable value after 7 ns. The simulated value of $M_z/M_s$ at 5 ns linearly increases with the external field, which does not agree with our experiment. However, the final magnetization state which is attained 2 ns after the end of current pulse agrees well with our results. (The domain structure and its analyses can be found in Supplementary S6 and S7.) The micromagnetic simulation including the damping-like torque closely resembles the behavior observed in the experiments, with substantially better agreement than previously reported single domain model calculations.

## Discussion

We designed a device based on the symmetric structure of Pt/CoNiCo/Pt to investigate the current switching of CoNiCo magnets by spin Hall effect from the bottom Pt layer. After excluding the thermal effect contribution, the spin Hall effect generates a damping-like effective field from the critical current density measurements of about 30 Oe per $10^7$ Acm$^{-2}$ with perpendicular magnetic fields, which is quite consistent with the results from the harmonic measurements (25 Oe per $10^7$ Acm$^{-2}$). No measurable $H_{FL}$, which originates mostly from the Rashba effect, was found in the device. In this study, comparing the switching abilities with the work done in Pt/Co/AlO$_x$ with Rashba field as high as several hundreds of Oersted in ref.

24, the switching effective field of symmetric Pt/FM/Pt is not decreased even without $H_{FL}$. Using the symmetric structure, the evidence is clear and direct compared with the asymmetric structure that the spin Hall effect is dominant for the current-induced switching in HM/FM structures.

We also performed the current-induced switching of the magnetization in the presence of in-plane magnetic fields. The critical current density is not sensitive to the magnitude and the direction of the in-plane magnetic fields. The deterministic switching of the ferromagnetic under in-plane fields was measured experimentally and then simulated by considering the damping-like torque due to the spin Hall effect. The current switching under in-plane fields is widely studied predominantly with the focus on the phase diagram of the switching critical current and the critical deterministic external fields[7,11] (the fields result in the full alignment of the magnetization in the z-direction). Below the critical fields, the relationship between $M_z$ and $H_x$ has been little studied. The results show non-linear dependence of $M_z$ on $H_x$ and the simulation predicted the same trend after considering the precession of the $M$ after the end of the current pulse. The works presented in this letter give us better understanding of the spin-orbital torque in HM/FM structure and applicable for the future design of spintronics devices.

## Methods

**Sample preparation.** Pt(3 nm)/CoNiCo(0.8 nm)/Pt(2 nm) thin films were sputtered onto Si/SiO$_2$ substrate at a base vacuum lower than $2 \times 10^{-5}$ Pa. Pt, Co and Ni layers were deposited at a working pressure of 0.5 Pa with the sputtering rate of 0.075 nm/s, 0.047 nm/s and 0.042 nm/s, respectively. Then, the top CoNiCo/Pt dots were first etched down to the bottom Pt/CoNiCo interfaces with diameter of 3 $\mu$m by E-beam lithography and ion milling. Using the same method, Hall bars were fabricated with the CoNiCo magnetic dot in the middle of the Hall cross.

**Electrical measurements.** The experiments use a Keithley 2602 current source and 2182 nano-voltmeter for the current switching measurements. The harmonic voltage measurements were conducted using two SR830 DSP lock-in amplifiers with the current frequency of 17 Hz.

Acknowledgements

This work was supported by "973 Program" No. 2014CB643903, 2015CB921401, and NSFC Grant Nos. 61225021, 11174272 , 11474272, 51431009 and 51471183 .

Author contributions

M.Y. and K. W. conceived the experiments. M.Y., K. C. and Y. S. fabricated the device. K. C. and M. Y. conducted the electrical measurements. H.J, S.L. and B.L. deposited the thin films. G. Y. and S. W. measured the FMR. M.Y. and K.C. conducted the micromagnetic simulations. Y. J. discussed the results. M.Y., K.C., K. W. E and K. W. wrote this paper.

Additional Information

The authors declare no competing financial interests.

Figures and figure Captions

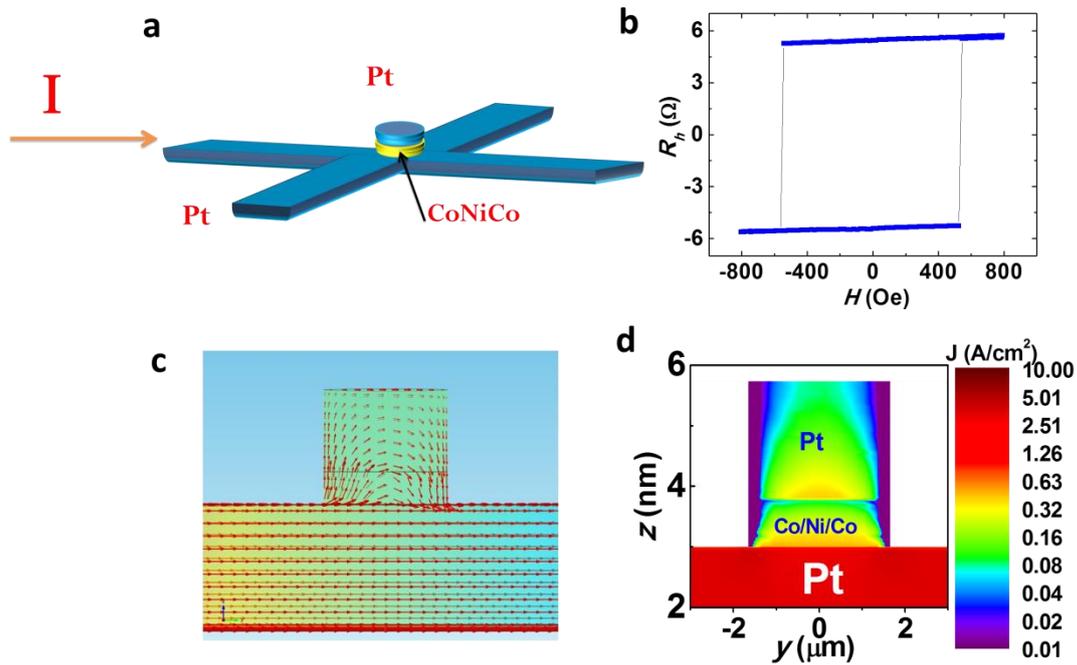

**Figure 1| The device structure and its current distribution.** (a) The structure of the device. The Hall cross in lateral direction is 8 μm wide which is used for applying the current. The transverse direction is 2 μm wide for the voltage detection. The CoNiCo/Pt dot is a circle with 3 μm in diameter. (b) The anomalous Hall resistance loop of the device under a perpendicular magnetic field. (c) The current flow route for the device obtained by finite element calculation. The orange background changing gradually to blue indicates the electric potential from high to low. (d) The current distribution projection along $x$ axis.

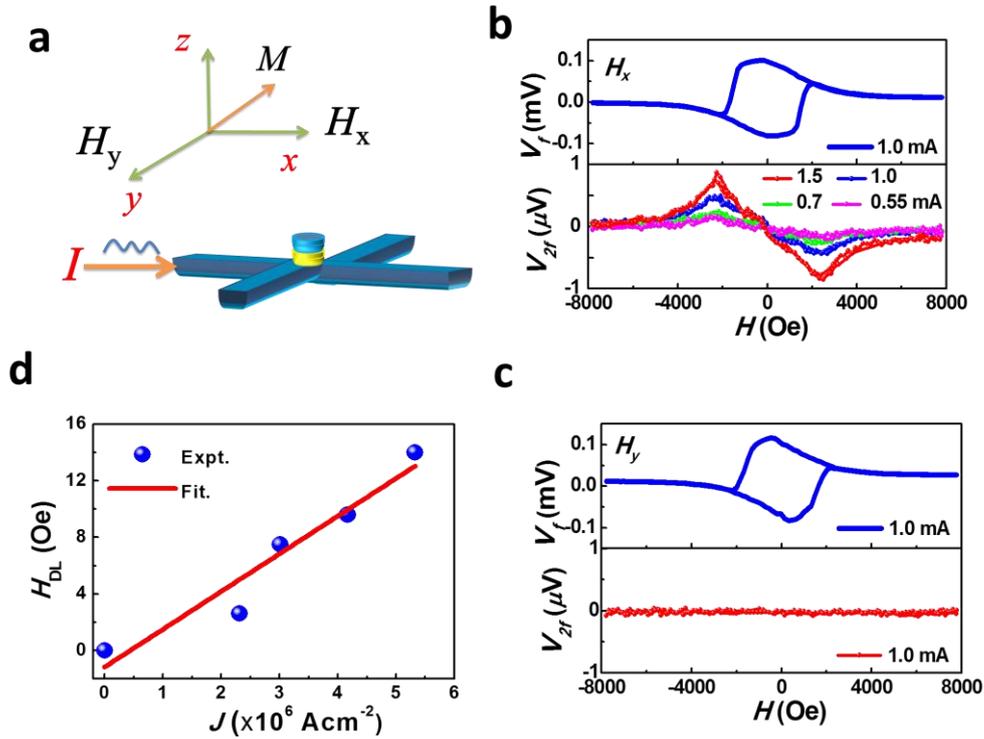

**Figure 2|** The effective fields of the Pt/CoNiCo/Pt device by harmonic measurements. (**a**) The measurement set up. (**b**) The first and second harmonic Hall voltage vs. the $H_x$ external field and (**c**) the $H_y$ external field. (**d**) The damping-like effective field calculated by the equation (1) and its linear fitting.

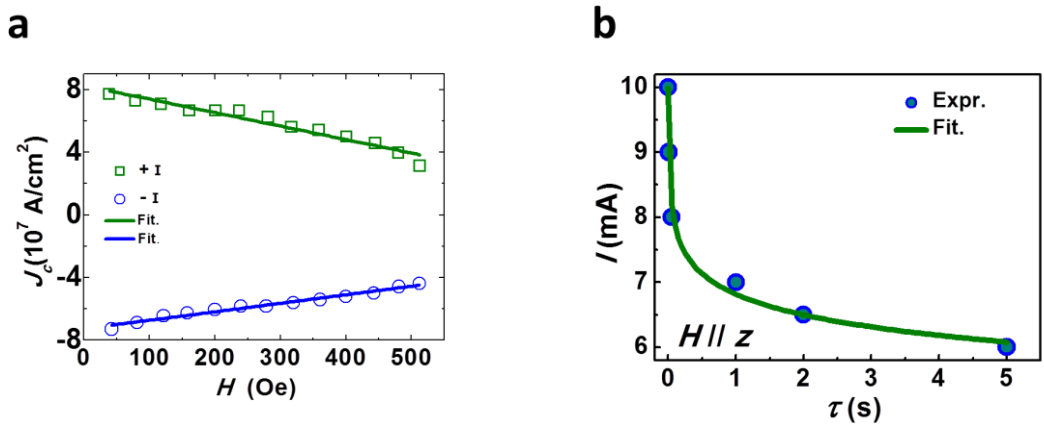

**Figure 3|** The effective fields of the Pt/CoNiCo/Pt device by switching current measurements. (**a**) The switching current density dependence on the perpendicular external field and its fitting using equation (2). (**b**) The critical switching current as a function of the duration of the current pulse and its fitting under the perpendicular field of 500 Oe. The $M$ of the magnet was first magnetized to $-M_z$ before the measurements.

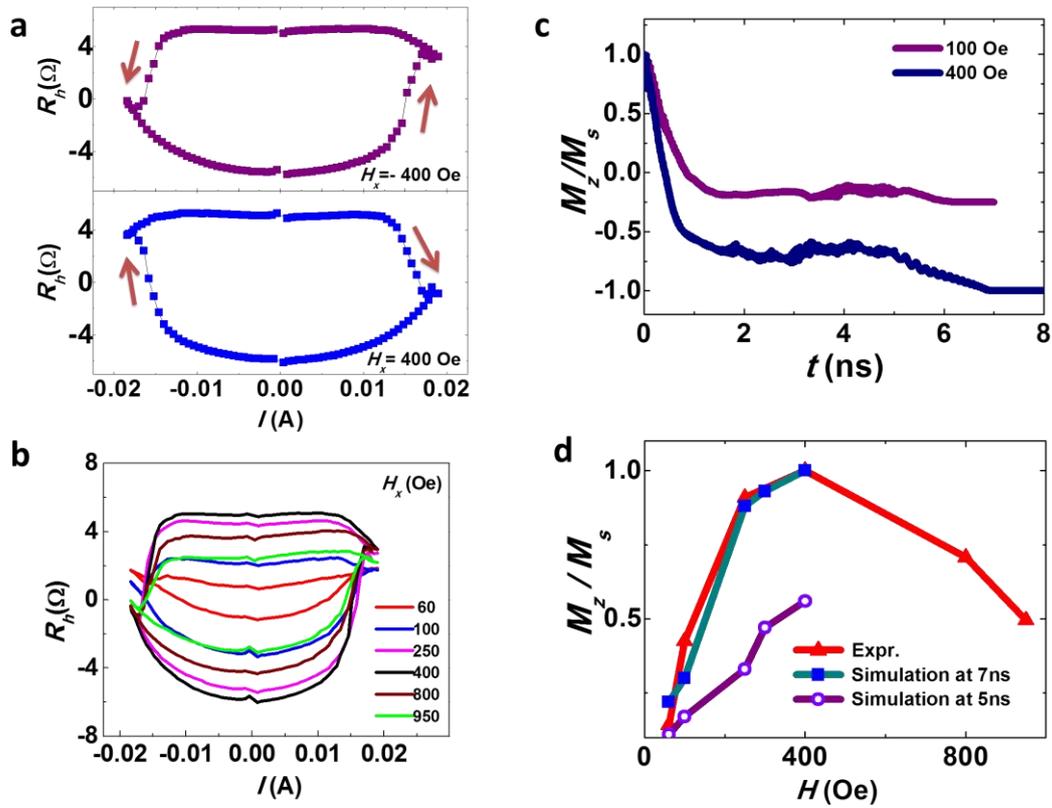

**Figure 4| Current switching under in-plane magnetic field**. (a) The current switching loop under an in-plane field of $\pm 400$ Oe in the *x*-direction, where the arrows indicate the current directions. (b) The current switching loop under different values of external $H_x$ field. (c) The simulation of $M_z/M_s$ for a 5 ns current pulse under external field of 100 Oe and 400 Oe. (d) The averaged $M_z$ after the current switching vs. the external $H_x$ field. The triangles represent the experimental data, the violet circles are the calculated values at the end of the 5 ns current pulse, and green squares are the final values 2 ns after the end of the current pulse.